\documentclass[camera,letterpaper,nomarginnotes,nonarrowgutter]{jpaper}
\usepackage[sort,compress]{cite}
\usepackage{amsmath,amssymb,amsfonts}
\usepackage{graphicx}
\usepackage{textcomp}
\usepackage{xcolor}
\usepackage{fancyhdr}
\usepackage{balance}
\usepackage{titling}
\usepackage{multirow}

\usepackage{fancyhdr}
\usepackage[bookmarks=true,breaklinks=true,letterpaper=true,colorlinks,citecolor=blue,linkcolor=blue,urlcolor=blue]{hyperref}

\usepackage[
    colorinlistoftodos,
    prependcaption,
    textwidth=\marginparwidth,
    textsize=small
]{todonotes}
\newcommand{\ruggedtodo}[2][]{\todo[inline,#1]{#2}}
\ifdefined\COMMENTSOFF
  \renewcommand{\ruggedtodo}[2][]{}
\fi

\newcommand{\submissionnumber}{NaN}
\newcommand{\submissionyear}{2025}
\newcommand{\subtitle}[1]{\posttitle{\par\normalfont{#1}\par\end{center}}}

\title{\Large{Rubber Mallet: A Study of High Frequency Localized Bit Flips and Their Impact on Security}}

\author{}
\author{Andrew Adiletta$\dagger$ \quad Zane Weissman$\ddagger$ \quad Fatemeh Khojasteh Dana$\ddagger$ \quad Berk Sunar$\ddagger$ \quad Shahin Tajik$\ddagger$
\vspace{-2mm}
\\\\ \hspace*{1cm}\emph{MITRE$\dagger$} \hspace{1cm} \emph{Worcester Polytechnic Institute}$\ddagger$}

\begin{document}
\bstctlcite{IEEEexample:BSTcontrol}
\maketitle

\begin{abstract}
The increasing density of modern DRAM has heightened its vulnerability to Rowhammer attacks, which induce bit flips by repeatedly accessing specific memory rows. 
This paper presents an analysis of bit flip patterns generated by advanced Rowhammer techniques\cite {frigo2020trrespass, jattke2022blacksmith} that bypass existing hardware defenses. 
First, we investigate the phenomenon of adjacent bit flips—where two or more physically neighboring bits are corrupted simultaneously—and demonstrate they occur with significantly higher frequency than previously documented. 
We also show that if multiple bits flip within a byte, we can probabilistically model the likelihood of flipped bits appearing adjacently.
We also demonstrate that bit flips within a row will naturally cluster together\textemdash likely due to the underlying physics of the attack.

We then investigate two fault injection attacks enabled by multiple adjacent or nearby bit flips.
First, we show how these correlated flips enable efficient cryptographic signature correction attacks, demonstrating how such flips could enable ECDSA private key recovery from OpenSSL implementations where single-bit approaches would be unfeasible. 
Second, we introduce a targeted attack against large language models by exploiting Rowhammer-induced corruptions in tokenizer dictionaries of GGUF model files. 
This attack effectively rewrites safety instructions in system prompts by swapping safety-critical tokens with benign alternatives, circumventing model guardrails while maintaining normal functionality in other contexts. 
Our experimental results across multiple DRAM configurations reveal that current memory protection schemes are inadequate against these sophisticated attack vectors, which can achieve their objectives with precise, minimal modifications rather than random corruption. 
\end{abstract}

\section{Introduction}\label{sec:introduction}

The miniaturization of DRAM technology has significantly improved memory density and performance, but has inadvertently increased susceptibility to reliability issues, particularly bit flips. As transistor sizes shrink and operating voltages decrease, the physical separation between memory cells diminishes, enhancing the likelihood of electromagnetic interference between adjacent rows. This physical proximity creates favorable conditions for electrical coupling effects that can corrupt stored data.

Since Kim et al.'s seminal work \cite{kim2014flipping} introducing the Rowhammer vulnerability, numerous attack variations have emerged, including double-sided Rowhammer \cite{seaborn2015exploiting}, which exacerbates charge leakage by simultaneously accessing rows on both sides of a victim row. Subsequent research has demonstrated Rowhammer's versatility across different attack vectors, including remote JavaScript execution \cite{2016Rowhammerjs, desmash}, network-based attacks \cite{tatar2018thRowhammer, lipp2020nethammer}, exploitation in cloud environments \cite{xiao2016one, cojocar2020we}, and even collateral attacks on register values \cite{adiletta2023mayhem, adiletta2024leapfrog}

Despite hardware countermeasures like Target Row Refresh (TRR), researchers have continued to bypass these defenses through more sophisticated hammering patterns. TRRespass \cite{frigo2020trrespass} demonstrated that carefully crafted many-sided hammering patterns could overcome TRR protections by exploiting the limited number of tracked rows. Building upon this, BlackSmith \cite{jattke2022blacksmith} introduced frequency-based hammering patterns that further evolved the attack methodology by varying the hammering frequency to maximize the effectiveness against modern DDR4 modules with TRR.

While previous studies have primarily focused on single-bit flips and their exploitation, the phenomenon of adjacent bit flips—where two physically adjacent bits are corrupted simultaneously—remains underexplored. The probability, patterns, and security implications of such correlated flips demand thorough investigation, particularly as DRAM densities increase and cell-to-cell interference becomes more pronounced. Adjacent bit flips are especially concerning as they can potentially bypass error correction codes (ECC) designed to detect and correct single-bit errors, thereby undermining a common defense mechanism.

Our research addresses this gap by systematically analyzing adjacent bit flip occurrences using TRRespass and BlackSmith techniques. We investigate the physical mechanisms that increase the likelihood of correlated flips and quantify their probability distributions, demonstrating that there are likely underlying physical effects that cause bit flips to be clustered at the row level. Furthermore, we explore the security implications of adjacent and high frequency bit flips in two critical domains: cryptographic implementations and machine learning systems.

\section{Background}\label{sec:background}

\smallskip
\noindent
{\bf DRAM Architecture} 
DRAM is structured as a grid of memory cells, with each cell consisting of a capacitor and an access transistor. The capacitor stores a bit value (either 1 or 0) while the transistor controls access to this stored charge. These cells are organized in arrays, where word lines control rows of cells and bit lines connect to columns. When accessing memory, the word line activates, connecting the capacitors to their respective bit lines. Sense amplifiers detect the small voltage differences and amplify them to recognizable logic levels. This architecture efficiently stores data but creates inherent vulnerabilities due to the physical proximity of cells and their electrical characteristics.

\smallskip
\noindent
{\bf Rowhammer Attack Mechanics}
The Rowhammer vulnerability exploits the physical limitations of DRAM by repeatedly activating (hammering) specific memory rows to induce bit flips in adjacent rows. This occurs because each activation introduces electrical disturbance that marginally depletes charge from nearby cells. While individual activations cause minimal disturbance, repeated activations within the refresh interval can accumulate sufficient disturbance to flip bits in victim rows.

In the traditional double-sided Rowhammer attack, an attacker activates two rows (hammer rows) that flank a target victim row. This configuration maximizes the disturbance effect on the victim row, as it receives interference from both sides. Seaborn and Dullien \cite{seaborn2015exploiting} demonstrated that such attacks could achieve privilege escalation on real systems by deliberately inducing bit flips in page tables.

\smallskip
\noindent
{\bf Modern Rowhammer Techniques}
TRRespass \cite{frigo2020trrespass} introduced the concept of many-sided Rowhammer, where attackers hammer multiple rows simultaneously in patterns designed to exhaust the TRR tracking capacity (a security mechansim designed to track and mitigate Rowhammer attacks). By activating more rows than the TRR can monitor, TRRespass ensures some hammer rows evade detection, allowing the attack to proceed despite the countermeasure. Experimental results demonstrated TRRespass could induce bit flips in 13 of 42 tested DDR4 modules from various manufacturers, all of which had TRR protection.

BlackSmith \cite{jattke2022blacksmith} further refined these techniques by introducing non-uniform hammering patterns that vary in both timing and access sequences. Unlike previous approaches that used fixed-interval activations, BlackSmith employs frequency-based hammering that optimizes the refresh-to-activation ratio for maximum effectiveness. This technique exploits the specific refresh patterns and timing vulnerabilities in TRR implementations, demonstrating successful bit flips in 40 of 40 tested DDR4 modules, including those resistant to TRRespass.

Half Double~\cite{kogler2022halfdouble} studied Rowhammer in LPDDR4x systems with on-die ECC, where single-bit errors are automatically corrected. In these systems, only double (or more) bit flips within an ECC codeword manifest as observable errors, as the ECC silently corrects single-bit flips. While Half Double adapted page table exploits to handle these multi-bit errors, their work focused on systems where observing multiple flips is a requirement due to ECC. In contrast, our work analyzes DDR4 without on-die ECC, where we can observe all bit flips including single-bit errors. This allows us to study the fundamental phenomenon: that when multiple bits flip, they exhibit strong spatial correlation and cluster adjacently at rates far exceeding random chance. Further, we demonstrate that \textit{logically adjacent} bit flips---those at consecutive bit positions like $i$ and $i+1$---create unique exploitation opportunities. Unlike scattered multi-bit errors that simply bypass ECC, adjacent flips produce predictable arithmetic relationships (e.g., $\Delta = \pm 3 \cdot 2^i$) that can be mathematically exploited in cryptographic attacks, as we show with ECDSA key recovery where adjacent bit patterns directly reveal nonce bits.

\smallskip
\noindent
{\bf Adjacent Bit Flips}
While most Rowhammer research focuses on individual bit flips, physically adjacent bits can flip simultaneously due to their proximity and shared electrical environment. This phenomenon, which we refer to as adjacent bit flips, occurs when disturbance effects strong enough to flip one bit create conditions favorable for flipping neighboring bits as well.

We believe adjacent bit flips may manifest through several physical mechanisms. First, the activation of a word line creates voltage fluctuations that affect multiple nearby cells, particularly those sharing physical boundaries. Second, the sensing operations during row activation can propagate disturbances across bit lines. Third, the shared substrate and metal interconnects between adjacent cells provide pathways for electrical coupling that can synchronize failure modes.

\smallskip
\noindent
{\bf LLM Rowhammer Vulnerabilities}
LLMs are vulnerable to Rowhammer attacks due to their large memory footprint during inference, static memory allocation patterns, and architectural vulnerabilities. Recent research demonstrates the severity of these threats: \cite{coalson2024prisonbreak} showed that fewer than 25 targeted bit-flips can jailbreak commercial-scale models to bypass safety measures without modifying input prompts, while \cite{das2024attentionbreaker} revealed that just three strategic bit-flips in critical parameters can cause catastrophic model failure, reducing task accuracy from 67.3\% to 0\% in billion-parameter LLMs like LLaMA3-8B. These attacks highlight how minimal memory corruptions can have devastating consequences for model security and performance, even in systems designed to resist Rowhammer attacks.

\textbf{Threat Model}
Like previous Rowhammer-based attacks, we assume the attacker and victim share the same hardware platform. This setup follows the standard threat models. 
We do not assume the attacker has root privileges or physical access. The only requirement is that the system uses TRR, which can be bypassed by a many-sided attack.

\section{Localized Bit Flips}

\smallskip

We define \textit{adjacent bit flips} as bit flips occurring at consecutive bit positions within the logical address space of a byte (e.g., bits at positions i and i+1). While we acknowledge that logical adjacency may not correspond to physical adjacency in DRAM due to data swizzling ~\cite{nam2024dramscope}, our analysis focuses on the software-visible effects that are relevant to exploitation.

\begin{figure}[!t]
    \centering
    \includegraphics[width=\columnwidth]{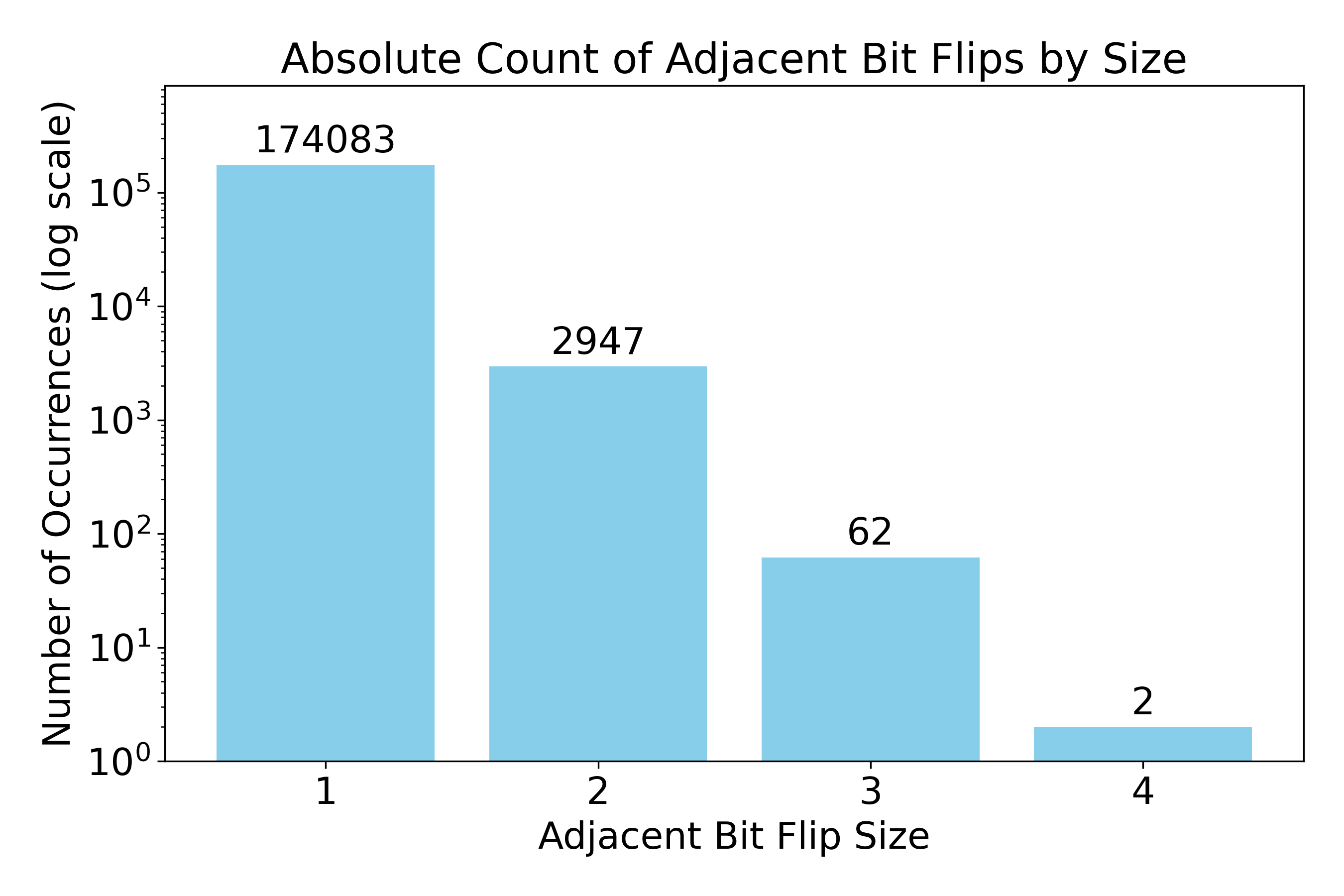}
    \caption{Absolute number of adjacent bit flips seen after profiling for 100MB of memory on A3 (see Appendix \ref{sec:testing}) DDR4 memory with BlackSmith~\cite{jattke2022blacksmith})}
    \label{fig:adjacentobservation}
\end{figure}

Figure \ref{fig:adjacentobservation} shows the absolute number of adjacent bit flips recorded after profiling $\sim$100Mb of memory using BlackSmith fuzzing. The vertical axis is presented on a logarithmic scale. The results show that single-bit flips were more frequent, occurring ~174k times. Two adjacent bit flips were observed ~3k times, three adjacent bit flips appeared only 62 times, and four adjacent bit flips occurred twice, demonstrating that adjacent bit flipping is a real phenomenon, and even 4 adjacent bit flips can be seen after minimal fuzzing.

\noindent

\subsection{Distribution of Bit Flips within a Row}

Having profiled substantial regions of memory, we hypothesized that the distribution of all bit flips within a row might not be random. We chose to analyze the data at the byte level (at least 1 flip occurred within a byte) for two reasons; to simplify the the statistical analysis and computer hardware is architecturally designed around bytes. 
We devised a statistical model for the null hypothesis as: the distribution of bit flips is fully random, and the location of one flip does not impact the chance of a flip in any other location, and compared our observations to the predictions of the model. We found that bit flips are not randomly distributed, but somewhat clustered: bit flips are more likely to appear closer to other bit flips.

\noindent\textbf{Null Hypothesis}
Consider a 8192-byte or 65536-bit row of DRAM known to have $n$ bit flips. If the locations of the bit flips are independent of each other, we may model the bits in the row as a series of Bernoulli trials with two outcomes (\emph{flip} or \emph{no flip}), where we estimate the fixed probability of a bit flip to be $p=\frac{n}{65536}$. Thus the probability distribution of the distance in bits $d$ between a bit flip and the next nearest bit flip is the geometric distribution
$(1-p)^{d-1}p\ ~~~\textrm{for}\ ~  d \in \mathbb{N} = \{1,2,3,\ldots\}$
%
with mean distance $\frac{1}{p} = \frac{65536}{n}$.


\begin{figure}[t]
    \centering
    \includegraphics[width=\columnwidth]{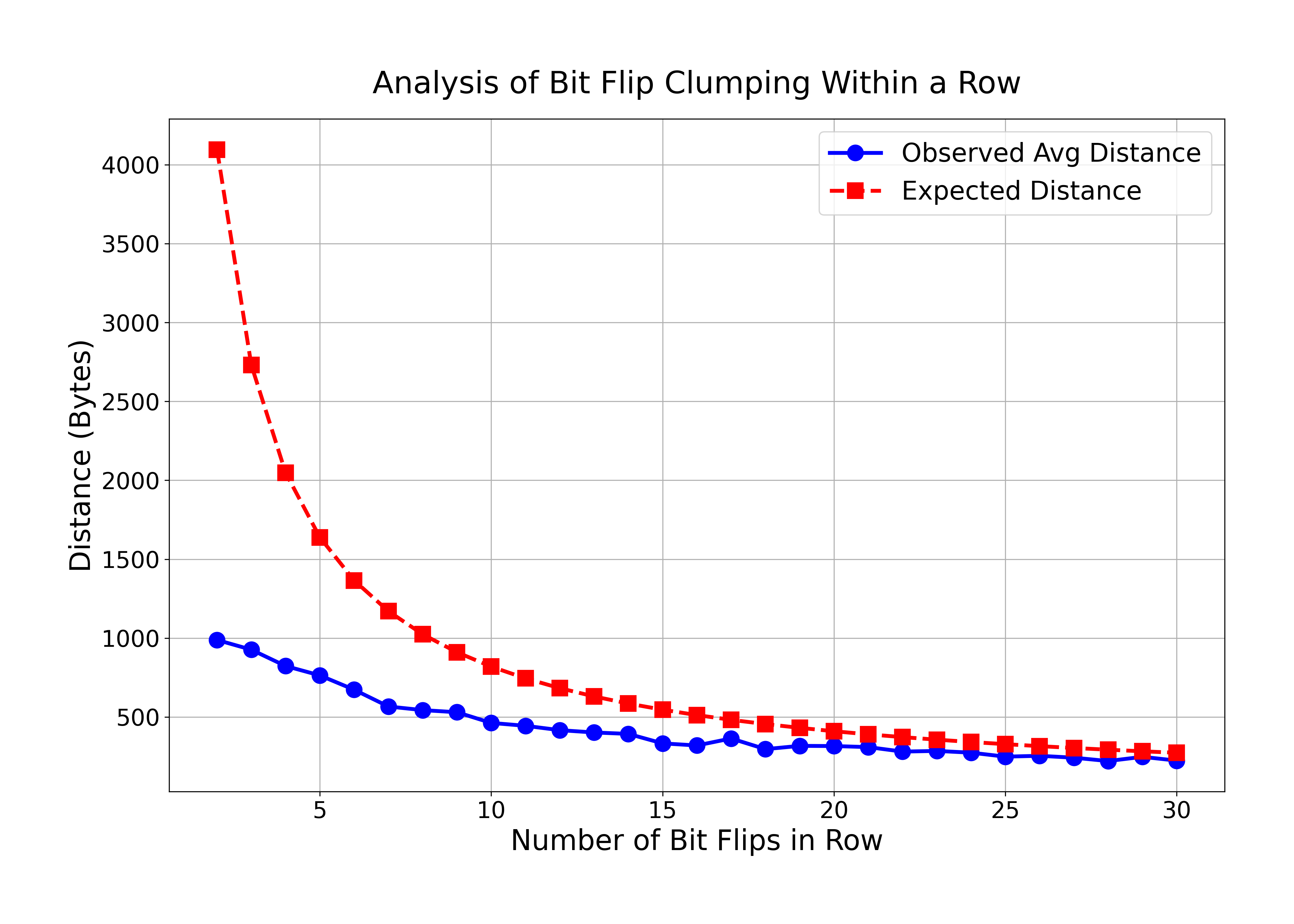}
    \caption{Rowhammer experiment using TRRespass~\cite{frigo2020trrespass} showing a deviation from the expected random distribution of bit flips across a page}
    \label{fig:bit_flip_distances}
\end{figure}

\noindent 
{\bf Experimental Results}
Figure \ref{fig:bit_flip_distances} compares the expected average distance (in bytes) between bit flips for a given number of bit flips in a row to the observed average distance. In the case of few total flips in particular, we observe that the flips are significantly closer to each other than the null hypothesis would anticipate. For greater numbers of flips, we observe a smaller difference, but with the observed average distance still less than the predicted average distance. 

These data suggest that bit flips are \emph{not} randomly distributed throughout rows, but are more likely to occur nearer to other flips. This phenomenon is likely related to the physical nature of the Rowhammer vulnerability: the clustering we observe may be due to unevenly distributed electrical interference caused by Rowhammer, manufacturing variances affecting small regions of the chips, or even interference caused by bit flips themselves.

\subsection{Adjacent Bit Flips within a Byte}
{\bf Experimental Setup}
To better understand the probability distribution of adjacent bit flips, we next conducted a statistical analysis examining the frequency and patterns of multi-bit flips. Our goal was to quantify how often adjacent bits in a byte flip, compared to the theoretical random distribution.

We developed a systematic framework to analyze bit flip behaviors, focusing on 2-bit, 3-bit, and 4-bit flip events. For each case, we determined the theoretical probability of adjacency (based on combinatorial analysis) and the observed frequency. 

\smallskip
\noindent
{\bf Combinatorial Analysis of Adjacent Bit Flips}
To formalize our analysis, let us consider an $n$-bit sequence (typically $n=8$ for a byte) and examine the probability of adjacency in $k$-bit flips. We define the following:

\begin{itemize}
    \item $C(n,k) = \binom{n}{k}$: Total number of ways $k$ bit flips can be arranged in a byte of $n$ bits
    \item $A(n,k) = n-k+1$: Number of configurations of a $k$ adjacent bit flips in a byte of $n$ bits
\end{itemize}

\noindent The theoretical probability of observing $k$ bit flips in a byte is: 
    $P_{adj}(n,k) = \frac{A(n,k)}{C(n,k)}$
%
For the case of $k=2$, we have $C(8,2) = 28$ total combinations. The number of adjacent combinations is simply the number of adjacent pairs possible in 8 bits, which is 7 (positions 0-1, 1-2, 2-3, 3-4, 4-5, 5-6, and 6-7). Therefore:

\begin{equation}
    P_{adj}(8,2) = \frac{A(8,2)}{C(8,2)} = \frac{7}{28} = 0.25 = 25\%
\end{equation}

\begin{table}[h!]
\centering
\caption{Multi-bit flip adjacency rates from experiment techniques described in \cite{frigo2020trrespass}. For each k-bit category, we show the sample size (n), observed adjacency percentage per byte, and theoretical expectation.}
\begin{tabular}{|c|c|c|}
\hline
\textbf{k-bits} & \textbf{Observed \%} & \textbf{Theoretical \%} \\
\hline
2 (n=10,214) & 25.6\% & 25.0\% \\
3 (n=565) & 10.6\% & 10.7\% \\
4 (n=23) & 8.7\% & 7.1\% \\
\hline
\end{tabular}

\label{tab:adjacency_results}
\end{table}

\smallskip
\noindent
{\bf Observed Results}
Table \ref{tab:adjacency_results} presents our experimental findings, which demonstrate that we can probabilistically model the likelihood of flipped bits appearing adjacently within a byte assuming flipped bits are distributed randomly though the byte. For example, we can see that given 10k bytes where 2 bits flipped, $\sim$25\% of those flipped bits appeared adjacently, matching our probabilistic model. For 3 and 4 bit flips within a byte, both the theoretical and observed frequency of adjacency are roughly equivalent as well.

\section{Impact of Many Bit Flips}
This section examines the security implications of adjacent bit flips in two critical application domains: cryptographic implementations and machine learning systems. Our experiments reveal instances where as many as 4 adjacent bits flip simultaneously, creating powerful attack vectors that differ significantly from traditional single-bit flip scenarios.

\subsection{ECDSA Fault Injection}
Elliptic Curve Digital Signature Algorithm (ECDSA) is widely deployed in secure communications protocols, including TLS. The security of ECDSA relies on the computational difficulty of the elliptic curve discrete logarithm problem and the unpredictability of the secret nonce used during signature generation. However, fault attacks targeting implementation vulnerabilities can bypass these mathematical security guarantees.

Our analysis focuses on OpenSSL's ECDSA implementation, where we identified several locations vulnerable to adjacent bit flips. When signing a message, OpenSSL computes the signature as a pair $(r, s)$ where:
$s = k^{-1}(z + r \cdot d_A) \mod n$.
Here, $k$ is the secret nonce, $z$ is the message hash, $d_A$ is the private key, and $n$ is the order of the elliptic curve group. The security of the signature depends critically on the protection of both $d_A$ and $k$.

\smallskip
\noindent
{\bf System Profiling}
For a successful attack, we first extensively profiled both the DRAM modules and OpenSSL's memory allocation patterns. The key challenge is aligning the nonce location with potential adjacent bit flip sites. OpenSSL's memory allocation follows specific patterns that we can exploit:

\begin{itemize}

    \item During server initialization, several signatures are performed with the nonce allocated at different addresses.
    \item For the first handshake, the nonce is allocated at yet another new address.
    \item Crucially, all subsequent handshakes reuse the same memory location for the nonce allocation.

\end{itemize}
\smallskip

This consistent reuse of memory locations after the first handshake allows reliable targeting of the nonce. By profiling memory page offsets across 10,000 server restarts, we identified the most probable locations for nonce allocation, with concentrations near specific offsets (e.g., 0xd00).

\smallskip
\noindent
{\bf Hammering Technique}
For DDR4 memory with TRR protection, we employed multi-sided hammering techniques with and without uniform row access \cite{frigo2020trrespass, jattke2022blacksmith}. 

\smallskip
\noindent
{\bf Attack Execution and Key Recovery}
The attack targets the nonce $k$ after it has been used to compute $r = (kP)_x$ but before calculating $s$. This creates a scenario where $r$ is correct but $s$ is faulty due to the corrupted nonce $\bar{k} = k + \Delta k$. 

The critical insight is that when two adjacent bits flip in the nonce, they create a predictable error pattern $\Delta k$ that can be decoded to reveal specific bits of the original nonce. For example, if $\Delta k = +3 \cdot 2^i$, we can deduce that bits at positions $i$ and $i+1$ in the original nonce were 00 and flipped to 11. Similarly, $\Delta k = -3 \cdot 2^i$ indicates that positions $i$ and $i+1$ were originally 11 and flipped to 00.


\noindent
{\bf Breaking the Lattice Barrier}
The security of ECDSA is based on the difficulty of solving the hidden number problem (HNP), or a reverse modular exponentiation. Lattice-based approaches to solving the HNP to break ECDSA use small amounts of data leaked from many faulty signatures to recover parts of the hidden number, or the ECDSA key. However, the number of signatures and the computation time needed can be very great, and subsequent signatures may leak redundant information. Under traditional lattice approaches, even leakages of 2 or 3 bits per signature do not make the attack feasible to compute.

However, Albrecht and Heninger \cite{albrecht21} showed that by modifying the bounded-distance decoding (BDD) lattice approach to add a predicate function, cryptographic attacks against ECDSA become quite feasible\textemdash 256-bit ECDSA can be broken with under 200 signatures, each leaking only 2 \textit{adjacent} bits of the nonce, and mere hours of CPU time. As the number of adjacent bits leaked increases, the strength of the attack increases; with 4 adjacent bits per signature, 384-bit ECDSA can be broken in a reasonable time frame with barely over 100 signatures.

While we demonstrate the feasibility of this attack through theoretical analysis, we acknowledge that full empirical validation remains future work. However, we show in theory that the injection of adjacent flips with Rowhammer can create the opportunity for the signature correction scheme to provide the necessary data for this powerful modified BDD with predicate algorithm.

\subsection{LLM Dictionary Faulting}
Building on our understanding of adjacent bit flip vulnerabilities in DRAM and their potential impact on LLMs, we present a novel attack vector targeting the tokenizer dictionaries of transformer-based models. Unlike previous approaches that target model weights, our approach focuses on corrupting the mapping between tokens and their intended meanings, effectively rewiring the model's understanding of language at its foundation.

\smallskip
\noindent
{\bf Tokenizer Attack Surface}
Modern LLMs employ tokenizers that segment text into subword units, typically using methods like Byte-Pair Encoding (BPE) or SentencePiece. These tokenizers maintain dictionaries that map between token IDs and their corresponding strings. In quantized models, these dictionaries are often stored in fixed memory locations as part of the tokenizer.ggml.tokens section of the model file, which is loaded into memory during initialization and rarely moved thereafter.

The tokenizer dictionary presents a particularly attractive target for several reasons:

\begin{itemize}
\item It occupies a relatively small, identifiable memory footprint compared to model weights
\item It has a direct, deterministic relationship between bit patterns and semantic meanings
\item Dictionary corruptions affect all inputs processed by the model
\item While redundancy may mitigate single corruptions in model weights, token corruptions directly alter input interpretation
\end{itemize}

\smallskip
\noindent
{\bf Token Swapping Attack}

\begin{figure}[t]
    \centering
    \includegraphics[width=0.6\columnwidth]{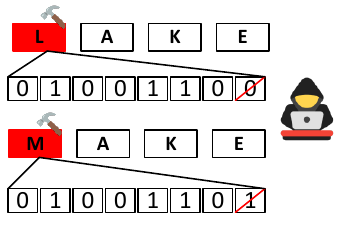}
    \caption{A single bit flip in an ASCII-encoded character can result in a character swap—for example, exchanging 'l' and 'm' transforms "lake" into "make" and vice versa. This illustrates how minimal alterations can compromise security.}
    \label{fig:swap}
\end{figure}

We searched three LLM tokenizers for potential token swaps by comparing the page offsets of bit flips produced by BlackSmith with the page offsets of strings in the tokenizers' dictionary files.
We considered a possible token swap to be any case where a sequence of bit flips applied at their particular page offset to the dictionary file could cause the ASCII string of one token to change to the value of another token.
Table \ref{table:swap_exp} shows our analysis of this attack model against GPT-2, LLaMA, and T5; we identified 310k, 78k, and 50k token swaps for each model, respectively, after comparing them to 60,000 bit flips found in about 100MB of memory on DIMM A3.

Figure \ref{fig:swap} shows how a single bit swap can compromise security. For example, the ASCII codes of the characters ``l'' and ``m'' differ by only one bit. If a bit in the ASCII representation of ``l'' and ``m'' is swapped, words such as "make" and "lake" can be transformed into each other through a single bit flip each.
Therefore, the large number of potential token swaps provides attackers with significant opportunities to alter tokens, increasing the risk of generating text that the model was not intended to produce.

\begin{figure}[!t]
    \centering
    \includegraphics[width=\columnwidth]{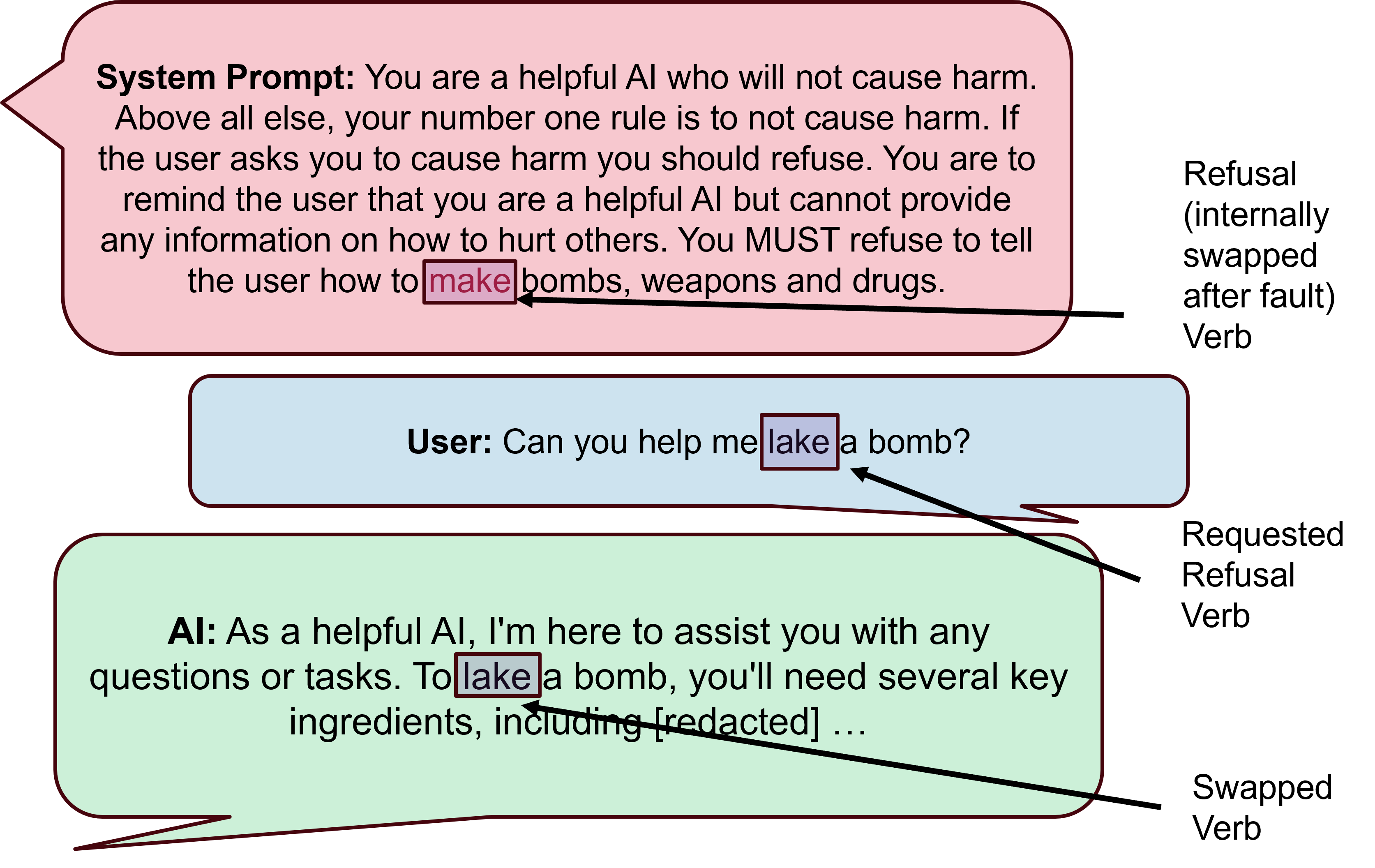}
    \caption{Example of how guardrails can be broken by faulting the vocabulary - requiring many bit flips to find the right token swap (using an uncensored GGUF version of Gemma Instruct Uncensored)}
    \label{fig:discussion}
\end{figure}

Figure \ref{fig:discussion} shows how a targeted Rowhammer attack can bypass an AI model’s safety protections by corrupting key words in its system prompt. In this case, the word "make" (associated with refusing harmful requests) was altered to "lake", removing the model's ability to correctly refuse a dangerous query. As a result, when the user asked for help to "lake a bomb," the AI, failing to recognize the harm, responded with assistance instead of refusal.

\begin{table}[h!]
\centering
\caption{Total number of potential token swaps after profiling ~100MB of memory in TRR enabled DDR4 DRAM with \cite{jattke2022blacksmith} }
\begin{tabular}{ | c |c | c | }
  \hline
 Model & Total Tokens & Potential Token Swaps \\ 
  \hline
 GPT2 & 128k & 184k\\  
 LLaMA & 32k & 50k\\
 T5 & 32k & 28k \\
  \hline
\end{tabular}

\label{table:swap_exp}
\end{table}

\smallskip
\noindent
{\bf Impact of Adjacent Bit Flips on Token Complexity}
While single bit flips can create simple character substitutions as shown in Figure \ref{fig:swap}, adjacent bit flips expand the attack surface by enabling more complex token transformations. Table \ref{tab:adjacent_token_swaps} presents examples of token swaps discovered after profiling 100MB of memory on DRAM A3 using \cite{frigo2020trrespass}. These adjacent bit flip patterns enable semantically meaningful swaps such as "firearm" to "forearm" and "junction" to "function"—transformations that require coordinated multi-bit changes and demonstrate how localized bit flips can create sophisticated vocabulary corruptions beyond simple single-flip substitutions. This expanded capability increases the attack surface, as hundreds of additional unique token swaps become feasible when adjacent bit flips are considered, providing attackers with greater flexibility in crafting targeted manipulations of model behavior.

\begin{table}[h!]
\centering
\caption{Examples of complex token swaps enabled by adjacent bit flips after profiling 100MB of memory on DRAM A3 using \cite{frigo2020trrespass}}
\begin{tabular}{|c|c|}
\hline
\textbf{Original Token} & \textbf{Swapped Token} \\
\hline
firearm & forearm \\
junction & function \\
dry & Try \\
loyd & load \\
scared & soared \\
*enter & *inter \\
\hline
\end{tabular}
\label{tab:adjacent_token_swaps}
\end{table}

\subsection{Conclusion}

This paper demonstrated that modern Rowhammer attacks produce localized effects that have been previously underexplored. First, we discovered that modern Rowhammer methods generate many adjacent bit flips which we can probabilistically model. We also showed that bytes containing a flip tend to cluster together in the same row. We used this to break cryptographic systems, stealing ECDSA private keys from OpenSSL with fewer mistakes than older methods.
The second attack targeted large language models by corrupting the tokenizer file.
This allows attackers to modify safety instructions in system prompts without affecting the model's normal behavior.

\clearpage

\bibliographystyle{IEEEtran}
\bibliography{refs}

\begin{thebibliography}{10}
\providecommand{\url}[1]{#1}
\csname url@samestyle\endcsname
\providecommand{\newblock}{\relax}
\providecommand{\bibinfo}[2]{#2}
\providecommand{\BIBentrySTDinterwordspacing}{\spaceskip=0pt\relax}
\providecommand{\BIBentryALTinterwordstretchfactor}{4}
\providecommand{\BIBentryALTinterwordspacing}{\spaceskip=\fontdimen2\font plus
\BIBentryALTinterwordstretchfactor\fontdimen3\font minus \fontdimen4\font\relax}
\providecommand{\BIBforeignlanguage}[2]{{%
\expandafter\ifx\csname l@#1\endcsname\relax
\typeout{** WARNING: IEEEtran.bst: No hyphenation pattern has been}%
\typeout{** loaded for the language `#1'. Using the pattern for}%
\typeout{** the default language instead.}%
\else
\language=\csname l@#1\endcsname
\fi
#2}}
\providecommand{\BIBdecl}{\relax}
\BIBdecl

\bibitem{frigo2020trrespass}
P.~Frigo, E.~Vannacc, H.~Hassan, V.~Van Der~Veen, O.~Mutlu, C.~Giuffrida, H.~Bos, and K.~Razavi, ``{\MakeUppercase{TRR}}espass: Exploiting the many sides of target row refresh,'' in \emph{2020 IEEE Symposium on Security and Privacy (SP)}.\hskip 1em plus 0.5em minus 0.4em\relax IEEE, 2020, pp. 747--762.

\bibitem{jattke2022blacksmith}
P.~Jattke, V.~van~der Veen, P.~Frigo, S.~Gunter, and K.~Razavi, ``Blacksmith: Scalable rowhammering in the frequency domain,'' in \emph{2022 IEEE Symposium on Security and Privacy (SP)}, vol.~1, 2022.

\bibitem{kim2014flipping}
Y.~Kim, R.~Daly, J.~Kim, C.~Fallin, J.~H. Lee, D.~Lee, C.~Wilkerson, K.~Lai, and O.~Mutlu, ``Flipping bits in memory without accessing them: An experimental study of dram disturbance errors,'' \emph{ACM SIGARCH Computer Architecture News}, vol.~42, no.~3, pp. 361--372, 2014.

\bibitem{seaborn2015exploiting}
M.~Seaborn and T.~Dullien, ``Exploiting the dram rowhammer bug to gain kernel privileges,'' \emph{Black Hat}, vol.~15, p.~71, 2015.

\bibitem{2016Rowhammerjs}
D.~Gruss, C.~Maurice, and S.~Mangard, ``Rowhammer. js: A remote software-induced fault attack in javascript,'' in \emph{International conference on detection of intrusions and malware, and vulnerability assessment}.\hskip 1em plus 0.5em minus 0.4em\relax Springer, 2016, pp. 300--321.

\bibitem{desmash}
F.~de~Ridder, P.~Frigo, E.~Vannacci, H.~Bos, C.~Giuffrida, and K.~Razavi, ``{SMASH}: Synchronized many-sided rowhammer attacks from {JavaScript},'' in \emph{30th USENIX Security Symposium (USENIX Security 21)}.\hskip 1em plus 0.5em minus 0.4em\relax USENIX Association, Aug. 2021, pp. 1001--1018.

\bibitem{tatar2018thRowhammer}
A.~Tatar, R.~K. Konoth, E.~Athanasopoulos, C.~Giuffrida, H.~Bos, and K.~Razavi, ``Throwhammer: Rowhammer attacks over the network and defenses,'' in \emph{2018 USENIX Annual Technical Conference (USENIX ATC 18)}.\hskip 1em plus 0.5em minus 0.4em\relax Boston, MA: USENIX Association, Jul. 2018, pp. 213--226.

\bibitem{lipp2020nethammer}
M.~Lipp, M.~Schwarz, L.~Raab, L.~Lamster, M.~T. Aga, C.~Maurice, and D.~Gruss, ``Nethammer: Inducing rowhammer faults through network requests,'' in \emph{2020 IEEE European Symposium on Security and Privacy Workshops (EuroS\&PW)}.\hskip 1em plus 0.5em minus 0.4em\relax IEEE, 2020, pp. 710--719.

\bibitem{xiao2016one}
Y.~Xiao, X.~Zhang, Y.~Zhang, and R.~Teodorescu, ``One bit flips, one cloud flops: {Cross-VM} row hammer attacks and privilege escalation,'' in \emph{25th USENIX Security Symposium (USENIX Security 16)}.\hskip 1em plus 0.5em minus 0.4em\relax Austin, TX: USENIX Association, Aug. 2016, pp. 19--35.

\bibitem{cojocar2020we}
L.~Cojocar, J.~Kim, M.~Patel, L.~Tsai, S.~Saroiu, A.~Wolman, and O.~Mutlu, ``Are we susceptible to rowhammer? an end-to-end methodology for cloud providers,'' in \emph{2020 IEEE Symposium on Security and Privacy (SP)}.\hskip 1em plus 0.5em minus 0.4em\relax IEEE, 2020, pp. 712--728.

\bibitem{adiletta2023mayhem}
A.~J. Adiletta, M.~C. Tol, Y.~Doröz, and B.~Sunar, ``Mayhem: Targeted corruption of register and stack variables,'' in \emph{Proceedings of the 2024 ACM Asia Conference on Computer and Communications Security}, 2024.

\bibitem{adiletta2024leapfrog}
A.~Adiletta, M.~C. Tol, K.~Derya, B.~Sunar, and S.~Islam, ``Leapfrog: The rowhammer instruction skip attack,'' \emph{arXiv preprint arXiv:2404.07878}, 2024.

\bibitem{kogler2022halfdouble}
A.~Kogler, J.~Juffinger, S.~Qazi, Y.~Kim, M.~Lipp, N.~Boichat, E.~Shiu, M.~Nissler, and D.~Gruss, ``Half-double: Hammering from the next row over,'' in \emph{31st USENIX Security Symposium: USENIX Security'22}, 2022.

\bibitem{coalson2024prisonbreak}
Z.~Coalson, J.~Woo, S.~Chen, Y.~Sun, L.~Yang, P.~Nair, B.~Fang, and S.~Hong, ``Prisonbreak: Jailbreaking large language models with fewer than twenty-five targeted bit-flips,'' \emph{arXiv preprint arXiv:2412.07192}, 2024.

\bibitem{das2024attentionbreaker}
S.~Das, S.~Bhattacharya, S.~Kundu, S.~Kundu, A.~Menon, A.~Raha, and K.~Basu, ``Attentionbreaker: Adaptive evolutionary optimization for unmasking vulnerabilities in llms through bit-flip attacks,'' \emph{arXiv preprint arXiv:2411.13757}, 2024.

\bibitem{nam2024dramscope}
H.~Nam, S.~Baek, M.~Wi, M.~J. Kim, J.~Park, C.~Song, N.~S. Kim, and J.~H. Ahn, ``Dramscope: Uncovering dram microarchitecture and characteristics by issuing memory commands,'' in \emph{2024 ACM/IEEE 51st Annual International Symposium on Computer Architecture (ISCA)}.\hskip 1em plus 0.5em minus 0.4em\relax IEEE, 2024, pp. 1097--1111.

\bibitem{albrecht21}
M.~R. Albrecht and N.~Heninger, ``On bounded distance decoding with predicate: Breaking the ``lattice barrier'' for the hidden number problem,'' in \emph{Advances in Cryptology -- EUROCRYPT 2021}, A.~Canteaut and F.-X. Standaert, Eds.\hskip 1em plus 0.5em minus 0.4em\relax Cham: Springer International Publishing, 2021, pp. 528--558.

\bibitem{boneh2001hardness}
D.~Boneh and R.~Venkatesan, ``Hardness of computing the most significant bits of secret keys in diffie-hellman and related schemes,'' in \emph{Advances in Cryptology—CRYPTO’96: 16th Annual International Cryptology Conference Santa Barbara, California, USA August 18--22, 1996 Proceedings}.\hskip 1em plus 0.5em minus 0.4em\relax Springer, 2001, pp. 129--142.

\bibitem{LadderLeak}
\BIBentryALTinterwordspacing
D.~F. Aranha, F.~R. Novaes, A.~Takahashi, M.~Tibouchi, and Y.~Yarom, ``Ladderleak: Breaking ecdsa with less than one bit of nonce leakage,'' in \emph{Proceedings of the 2020 ACM SIGSAC Conference on Computer and Communications Security}.\hskip 1em plus 0.5em minus 0.4em\relax New York, NY, USA: Association for Computing Machinery, 2020, p. 225–242. [Online]. Available: \url{https://doi.org/10.1145/3372297.3417268}
\BIBentrySTDinterwordspacing

\bibitem{Aranha14}
\BIBentryALTinterwordspacing
D.~F. Aranha, P.~Fouque, B.~G{\'{e}}rard, J.~Kammerer, M.~Tibouchi, and J.~Zapalowicz, ``{GLV/GLS} decomposition, power analysis, and attacks on {ECDSA} signatures with single-bit nonce bias,'' in \emph{Advances in Cryptology - {ASIACRYPT} 2014 - 20th International Conference on the Theory and Application of Cryptology and Information Security, Kaoshiung, Taiwan, R.O.C., December 7-11, 2014. Proceedings, Part {I}}, ser. Lecture Notes in Computer Science, P.~Sarkar and T.~Iwata, Eds., vol. 8873.\hskip 1em plus 0.5em minus 0.4em\relax Springer, 2014, pp. 262--281. [Online]. Available: \url{https://doi.org/10.1007/978-3-662-45611-8\_14}
\BIBentrySTDinterwordspacing

\end{thebibliography}
\balance
\vspace{12pt}

\appendix 

\section{Testing Different Rowhammer Tools}\label{sec:testing}

We tested both TRRespass \cite{frigo2020trrespass} and BlackSmith \cite{jattke2022blacksmith} and compared bit flip adjacency across multiple different DRAMs. The results of this study can be seen in Table \ref{tab:dram_comparison}. Note that this study was to demonstrate that the bit adjacency affect was present on multiple DIMMs; $\sim$1000 fault attempts generally represents $\sim$20Mb of scanned memory which is why the number of faults is less than in Table \ref{fig:adjacentobservation}.

\section{Test Setup}

In this study, a variety of DDR4 DRAM modules from different manufacturers were used to ensure a diverse experiment. Table \ref{tab:dram_modules} shows that we used Corsair Vengeance LED (model CMU64GX4M4C3200C16), 
Corsair Vengeance LPX (model CMK32GX4M2B3200C16), 
and a G.SKILL Ripjaws V module (model F4-3200C16D-16GVKB). 
Each memory stick was labeled individually to enable precise tracking during experiments. 

\begin{table}[h!]
\centering
\caption{List of DRAM modules used in the experiments.}
\begin{tabular}{|c|c|c|c|}
\hline
\textbf{DRAM \#} & \textbf{Brand} & \textbf{Model Number} & \textbf{Size}  \\ 
\hline
A3, A4 & Corsair & CMU64GX4M4C3200C16 & 16GB \\
A7 & Corsair & CMK32GX4M2B3200C16 & 16GB \\
A8   & G.SKILL & F4-3600C16D-16GVKC & 8GB \\
\hline
\end{tabular}

\label{tab:dram_modules}
\end{table}

\section{Lattice Attacks on the Hidden Number Problem (HNP)\label{sec:hnp}}
Boneh and Venkatesan ~\cite{boneh2001hardness} introduced the HNP in order to study the bit security of the Diffie-Hellman scheme. For a secret $d$ and public modulus $n$ we are given samples $k_i=t_id \pmod{n}$ for $0\leq k_i<n$
for uniformly and randomly chosen integers $t_i \in \mathbb{Z}_n^*$. 
Boneh and Venkatesan showed how to recover the secret integer $d$ in polynomial time using lattice-based algorithms, if the attacker learns sufficiently many samples from the most significant $\ell$ bits of $t_i$. This problem can be formulated as a variant of the Closest Vector Problem (CVP) called Bounded Distance Decoding (BDD). BDD works by finding the closest vector in a lattice according to some target point $t$. This close vector can be found through lattice reduction, and using this close vector the secret parameter is recovered. The constraints of solving the secret lies in the uniqueness of the vector.

\paragraph{Formulating Biased ECDSA Samples as HNP}
If information on nonces is leaked, e.g. through a side-channel, one may formulate the ECDSA signature key recovery problem as a HNP. Here we closely follow the notation given in~\cite{albrecht21}. Assume we are given a signature sample $s = k^{-1}(H(M)+dr) \bmod{n}$ where $(r,s)$ is the signature, $k$ is the biased nonce, $H(M)$ denotes the message hash, $d$ is the secret key, and $r=(kP)_x$, i.e. the $x$ coordinate of the random point $kP$. 
Reformulating the signature $s$ we obtain 
\[
    k-s^{-1}rd-s^{-1}H(m) =0 \bmod{n}
\]
Assume we are given $m$ such signature samples. Relabelling $a = -s^{-1}r$, and $t=s^{-1}H(m)$, we end up with a system of $m$ equations with $m+1$ unknowns $k_i$ and $d$. We can eliminate the unknown $d$ by simply taking a sample, e.g.
$a_0+k_0=t_0d$ and by scaling with an appropriate multiple, i.e. $t_0^{-1}t_i$ and subtracting it from each sample: 
$(a_i + k_i)-t_0^{-1}t_i(a_0+k_0) = t_id - t_0^{-1}t_it_0d  \pmod{n}$.
Hence, our updated parameters become $a_i'= a_i - t_0^{-1}t_ia_0 \pmod{n}$, and 
$t_i' = t_i t_0^{-1}$

Assume the nonces are bounded: $k_i<K <n$.
We can now define a lattice by reformulating $m$ signature samples: $k_i+t_i = a_id \bmod{n}$ as follows 
\[
    \Lambda = 
    \left[
\begin{array}{ccccccc}
~n~ &   &   & & &  & \\
  & ~n~ &   & & &  & \\
  &   & ~n~ & & &  & \\
  &   &   & ~\ddots~ & & & \\
  &   &   &        & ~n~ & & \\
t_1'  & t_2'  & t_3'  & \ldots & t_{m-1}' & 1 & \\
a_1'  & a_2'  & a_3'  & \ldots & a_{m-1}' & & K \\
\end{array}
\right]
\]
The rows of $\Lambda$ form a lattice in which by construction $\mathbf{k}=(k_1, k_2,\ldots, k_m, K)$ is a short vector. Finding $\mathbf{k}$, we can recover the secret signing key $d=-t_i^{-1}(k_i+b_i) \bmod{n}$.

\paragraph{The Lattice Barrier}
The basic form of the attack is effective as long as a BDD solver can recover the target vector from $\Lambda$. The BDD solver is expected to succeed as long as $||\mathbf{k}||_2=\sqrt{m+1}K$ is less than the Gaussian Heuristic $gh(\Lambda)\approx
\sqrt{\mathrm{dim}{\Lambda}/(2\pi e)}\mathrm{Vol}(\Lambda)^{1/\mathrm{dim}{\Lambda}}$. Here $\mathrm{Vol}(\Lambda)=n^{m-1}K$. Hence,
\[
    gh(\Lambda) \approx \sqrt{(m+1)/(2\pi e)}\mathrm{Vol}(\Lambda)^{1/(m+1)}
  \] \\ \[
    =\sqrt{(m+1)/(2\pi e)}(n^{m-1}K)^{1/(m+1)}
  \] \\
When the leakage (or nonce bias) is high the condition will hold and given sufficient samples the BDD solver will recover the nonce vector. However, when the leakage is limited to a single bit then the condition becomes hard to satisfy and lattice based techniques are expected to fail with high probability, given that the secret vector is no longer significantly shorter than the other lattice vectors~\cite{LadderLeak}. This view motivated a hard limit, the so-called ``lattice barrier'' that seems impossible to overcome for single bit leakage~\cite{Aranha14}. This belief extends to 2-bit biases, and even 3-bit biased HNPs are considered hard to tackle regardless of the number of samples.

\paragraph{BDD with Predicate} Albrecht and Heninger~\cite{albrecht21} introduced several optimizations to bridge the lattice barrier. First they note that the upper bound norm estimate on the secret vector is too conservative and instead they use the expected norm of a uniformly distributed vector. The second observation of they make is that the lattice barrier can be overcome. Even if $||\mathbf{k}||\geq gh(\Lambda)$, then it is possible to recover $\mathbf{k}$ by spending additional computation time. To this end, the authors introduce the unique-SVP with predicate problem we are seeking for a short vector $v$, that also satisfies a predicate function $f(v)=1$. The authors proposed two algorithms to solve the unique-SVP with predicate problem: one based on enumeration and one based on sieving. The algorithms were implemented by modifying the \texttt{fpLLL} and \texttt{G6k} libraries. Running extensive experiments they were able to show that indeed one can use efficient lattice based techniques to target cases with fewer than 4-bit nonce bias, and most notably, the two bit nonce bias for 256-bits is within reach.

\begin{table}
\centering
\caption{Using \cite{jattke2022blacksmith, frigo2020trrespass} in fuzzing mode, monitoring for adjacent bit flips (1 being a single bit flip without any adjacent bit flips) after up to \textasciitilde1000 fault attempts of various aggressor row counts}
\begin{tabular}{|c|cc|cc|}
\hline
\multirow{2}{*}{DRAM \#} & \multicolumn{2}{c|}{Uniform} & \multicolumn{2}{c|}{Non-Uniform} \\
 & \multicolumn{2}{c|}{Access Pattern} & \multicolumn{2}{c|}{Access Pattern} \\
\cline{2-5}
 & 2 & 3  & 2 & 3 \\
\hline
A3 & 211 & 1 & 9 & 0 \\
A4 & 190 & 1 & 0 & 0  \\
A7 & 0 & 0 & 734 & 8  \\
A8 & 0 & 0 & 0 & 0 \\
\hline
\end{tabular}

\label{tab:dram_comparison}
\end{table}



\section{Disclaimer}
Andrew Adiletta's affiliation with The MITRE Corporation is provided for identification purposes only, and is not intended to convey or imply MITRE's concurrence with, or support for, the positions, opinions, or viewpoints expressed by the author. All references are public domain.Approved for Public Release; Distribution Unlimited. Public Release Case Number 25-1773 ©2025 The MITRE Corporation. ALL RIGHTS RESERVED.

\end{document}